\documentstyle[12pt,
epsfig,
cite,%
amssymb,
a4p]{article}
\begin{document}
\textheight=23.cm

\date{}

\def\jour#1#2#3#4{{#1} {\bf#2} (19#3) #4}
\def\appj#1{{#1}}
\def\PL{Phys. Lett. {\bf B}}
\def\ZP{Z. Phys. {\bf C}}
\def\EPJ{Eur. Phys. J. {\bf C}}
\def\NP{Nucl. Phys. {\bf B}}
\def\PRp{Phys. Reports}
\def\PRD{Phys. Rev. {\bf D}}
\def\PRC{Phys. Rev. {\bf C}}
\def\IJ{Int. J. Mod. Phys. {\bf A}}
\def\ML{Mod. Phys. Lett. {\bf A}}
\def\JP{J. Phys. {\bf G}}
\def\AP{Acta Phys. Pol. {\bf B}}
\def\NIM{Nucl. Instr. Meth. {\bf A}}
\def\CP{Comp. Phys. Comm.}

\def\nwp{\newpage}
\def\bi{\bibitem}
\def\vs{\vspace*}
\def\hs{\hspace*}
\def\ct{\cite}
\def\be{\begin{equation}}
\def\ee{\end{equation}}
\def\bea{\begin{eqnarray}}
\def\eea{\end{eqnarray}}
\def\la{\label}
\def\bc{\begin{center}}
\def\ec{\end{center}}

\def\al{\langle}
\def\ar{\rangle}

\def\leq{\leqslant}
\def\geq{\geqslant}
\def\lssim{\stackrel{<}{_\sim}}
\def\gtsim{\stackrel{>}{_\sim}}

\def\ea{{\sl et al.}}
\def\eg{{\sl e.g.}}
\def\et{{\sl etc.}}
\def\ie{{\sl i.e.}}
\def\va{{\sl via }}
\def\vrs{{\sl vs. }}

\def\ep{e$^+$e$^-$ }
\def\z{Z$^0$}
\def\pT{p_T}
\def\phi{\Phi}
\def\yf{y$$\times$$\phi}
\def\yp{y$$\times$$\pT}
\def\fp{\phi$$\times$$\pT}
\def\3d{y$$\times$$\phi$$\times$$\pT}
\def\od{single-particle }

\def\err{uncertainties }
\def\errp{uncertainties}
\def\mode{in comparison with the predictions of  two \MC models.} 
\def\lides{The total error is shown along with the statistical error 
(inner error bars) for each point.
} 
\def\vliet{The error bars show the total \errp.} 
\def\int{intermittency }
\def\intp{intermittency}
\def\mom{moments }
\def\momp{moments}
\def\cum{cumulants }
\def\cump{cumulants}
\def\mupa{multiparticle }
\def\flus{fluctuations }
\def\cors{correlations }
\def\flup{fluctuations}
\def\corp{correlations}
\def\gen{genuine }
\def\phs{phase space}
\def\psh{phase-space}
\def\MC{Monte Carlo }
\def\HW{{\sc Herwig} }
\def\JT{{\sc Jetset} }
\def\OP{OPAL }
\def\DE{DELPHI }
\def\col{Collaboration}

\def\fig{Fig. }
\def\fgs{Figs. }
\def\rfl{Ref. }
\def\rfs{Refs. }
\def\frm{Eq. }
\def\fre{Eqs. }


\renewcommand{\Huge}{\huge}

\title{\vs{4.cm}
\bf Intermittency and Correlations\\ in Hadronic {\z}
Decays\\}

\author{\Large The OPAL Collaboration}
\maketitle

\thispagestyle{empty}

\vs{-9.7cm}
\begin{center}
{\Large EUROPEAN LABORATORY FOR PARTICLE PHYSICS}
\end{center}
\vs{.5cm}
\begin{flushright}
{\large CERN-EP/99-009\\
28th January 1999}
\end{flushright}

\vs{5.cm}


\begin{abstract}
\noindent

\noindent A  multidimensional study of local multiplicity \flus and
multiparticle \cors of hadrons produced in {\z} decays is
performed. The study is based on the data sample of more than
4$\times$10$^6$
events recorded with
the OPAL detector at LEP. The \flus and \cors are analysed in
terms of the normalized scaled factorial \mom and \cum up to the fifth
order.
The \mom are observed to have \intp-like behaviour,
which is found to be more pronounced with increasing dimension. The
large data sample allows for the first time a study of the factorial
\cum in \ep annihilation.
The analysis of the \cum shows the existence of genuine multiparticle
\cors with a strong \int rise up to higher orders. 
These \cors are found to be stronger in higher dimensions.
The decomposition of the factorial \mom into lower-order
\cors shows that the dynamical \flus have important contributions from
\gen many-particle \corp.
The \MC models \JT 7.4 and \HW 5.9 are found to
reproduce the trend of the measured moments and cumulants
but they underestimate the
magnitudes. The results are found to be consistent
with QCD jet formation dynamics, although additional
contributions from other mechanisms in the hadronization process cannot be
excluded.

\end{abstract}
\vs{1.2cm}
\centerline{\Large (Submitted to  European Physical Journal C)}

\nwp
\pagestyle{plain}
\setcounter{page}{1}


\begin{center}{
G.\thinspace Abbiendi$^{  2}$,
K.\thinspace Ackerstaff$^{  8}$,
G.\thinspace Alexander$^{ 23}$,
J.\thinspace Allison$^{ 16}$,
N.\thinspace Altekamp$^{  5}$,
K.J.\thinspace Anderson$^{  9}$,
S.\thinspace Anderson$^{ 12}$,
S.\thinspace Arcelli$^{ 17}$,
S.\thinspace Asai$^{ 24}$,
S.F.\thinspace Ashby$^{  1}$
% ...
}\end{center}\bigskip
\bigskip
$^{  1}$School of Physics and Astronomy, University of Birmingham,
Birmingham B15 2TT, UK
\newline
$^{  2}$Dipartimento di Fisica dell' Universit\`a di Bologna and INFN,
I-40126 Bologna, Italy
\newline
% ...
$^{  5}$Cavendish Laboratory, Cambridge CB3 0HE, UK
\newline
$^{  6}$Ottawa-Carleton Institute for Physics,
Department of Physics, Carleton University,
Ottawa, Ontario K1S 5B6, Canada
\newline
% ...
$^{  8}$CERN, European Organisation for Particle Physics,
CH-1211 Geneva 23, Switzerland
\newline
$^{  9}$Enrico Fermi Institute and Department of Physics,
University of Chicago, Chicago IL 60637, USA
\newline
% ...
$^{ 12}$Indiana University, Department of Physics,
Swain Hall West 117, Bloomington IN 47405, USA
\newline
% ...
$^{ 16}$Department of Physics, Schuster Laboratory, The University,
Manchester M13 9PL, UK
\newline
$^{ 17}$Department of Physics, University of Maryland,
College Park, MD 20742, USA
\newline
% ...
$^{ 23}$Department of Physics and Astronomy, Tel Aviv University,
Tel Aviv 69978, Israel
\newline
$^{ 24}$International Centre for Elementary Particle Physics and
Department of Physics, University of Tokyo, Tokyo 113-0033, and
Kobe University, Kobe 657-8501, Japan
\newline
% ...
\newline

\nwp


\nwp
\thispagestyle{empty}


\nwp

\section{Introduction}
\la{intro}

Particle density {\flus} of 
hadronic final states produced in high-energy 
collisions have been extensively investigated in the last decade.  For
recent reviews, see {\eg} \rfs \ct{rev1,rev2}.  Dynamical
({\ie}, non-statistical) \flus were observed,
establishing the phenomenon of \intp, the increase of
factorial moments with decreasing bin size \ct{bp}. 
The \int approach of studying the distributions of particles
in restricted regions of {\phs} allows a detailed
analysis of the dynamics of hadroproduction.
Furthermore, the behaviour of the factorial moments shows the
self-similar nature of density \flup, {\ie}, the particle distributions
show similar fluctuations on all resolution scales, a 
characteristic of fractals \ct{frac}.   
\\

Despite numerous experimental and theoretical studies, the origin of \int
remains unclear,
although important features of this effect have been observed \ct{rev2}.
For example, experimental investigations have shown an enhancement of the
phenomenon in \ep annihilation as compared to hadronic and nuclear
collisions. 
Furthermore, larger intermittency effects have been observed when
several dynamical variables are considered together, as compared to the
effect seen in one-dimensional analyses. 
Existing \MC (MC) models, which use parton shower simulations and
differing fragmentation and hadronization models,  simulate most
details
of \ep collisions and the general properties of hadronic interactions
well, but fall short in predicting the intermittent structure found in the
data, both in \ep collisions and in hadronic interactions. 
Theoretical approaches have not clarified the origin of the observed
dynamical \flup.  
The intermittent behaviour of particle distributions may prove to be a
strong test of QCD, which already provides guidelines for explaining the
``soft'' character of \int \ct{revd,revo}. 
\\

The goal of this study is to investigate the dynamical correlations
of many-particle systems produced in \ep annihilation.
One must be careful to separate out
the effects of lower-order correlations when searching for higher-order
ones.  For example, a correlation in the production of pairs of particles
in neighboring regions of {\phs} will necessarily induce correlations
when particles are considered three at a time.
It has been suggested that \int should therefore
be analysed in terms of the factorial
cumulant \mom to reveal ``genuine'' \mupa \cors by not being
sensitive to the contributions of lower-order correlations \ct{cum}. 
The
investigations carried out for heavy-ion reactions have not shown any
\cors higher than two-particle ones, while in studies of hadron-hadron
collisions significant higher-order \cors have been observed,
although the latter have been seen to weaken with increasing multiplicity
at a fixed centre-of-mass energy \ct{rev2}.  These effects 
have been explained as a consequence of
the events consisting of superpositions of multiple independent particle
sources \ct{isc}. These findings suggest that interactions with a low
number of very hard scattering processes, such as high-energy \ep
annihilation,    
might be more sensitive to \gen \mupa \corp.
\\

This paper describes the study of the \int phenomena and the \gen \mupa
\cors of charged particles in the three-dimensional {\phs} of rapidity,
transverse momentum, and azimuthal angle, as defined in Section 
\ref{data}. 
This analysis uses more than four million multihadronic events recorded by
the \OP detector at the LEP collider with $\sqrt{s}\approx m_{\rm Z^0}$. 
This data sample is much larger than that used in OPAL's previous
publication \ct{O}, and in other \ep investigations carried out at the
{\z} peak \ct{D,A,L3,291,SL} and at lower energies \ct{291,le}.  The
statistical precision of this data sample allows us
to extend the
former intermittency analysis \ct{O} to a multidimensional one with the
possibility to reach high-order \flus at very small bins. 
With this high statistics
we are able for the first time to search for \gen \mupa \cors
by means of normalized factorial cumulants in \ep collisions.  \\

The paper is organised as follows. In Section \ref{method} the normalized
factorial \mom and factorial cumulant \mom are introduced and the
technique of extracting dynamical \flus and \gen \mupa \cors is given. 
The detector, data sample and  correction procedure are
described in Section \ref{data}. The results and their comparison
with MC predictions are presented and discussed in Section \ref{res}.
In Section \ref{sum} both a summary and conclusions are presented.

\section{The method}
\la{method}

\subsection{Scaled factorial \mom and \int}
\la{fmi}

To search for local dynamical \flus we use the 
normalized 
scaled factorial \mom introduced in \rfl \ct{bp}. 
The \mom are defined as

\be
F_q =
{\cal N}^q
\left\al \overline {n_m^{[q]}}\right\ar /\bar N_m^{[q]}.
\la{fmh}
\ee
Here $n_m^{[q]}$ is the $q$th order factorial multinomial,
$n_m(n_m-1)\cdots(n_m-q+1)$,
with $n_m$ particles in the $m$th bin of the {\phs} ({\eg} rapidity
interval) 
divided into $M$ equal bins. $N_m$ is the number of particles in the $m$th
bin summed over all the $\cal N$ events in the sample,
$N_m= \sum_{j=1}^{\cal N}(n_m)_j$.
The bar indicates averaging over the bins in each event,
$(1/M)\sum_{m=1}^{M}$ (``horizontal'' averaging), while the angle
brackets denote averaging over the events (``vertical'' averaging).\\

The \mom in \frm (\ref{fmh}) are given in the  modified
form, in contrast 
to those used in the earlier \ep studies \ct{O,D,A,L3,291,SL,le}. This
form
has been
suggested in \rfl \ct{ks} to take into account the bias arising 
from the assumption of infinite statistics
in the normalization calculation \ct{bp,pwhy} and affects the \momp,
particularly those computed 
with small bins.  
\\

These moments, defined in \frm (\ref{fmh}), are the so-called
``horizontal'' \mom \ct{pwhy}, where the \flus are first averaged over all
$M$ bins in each event and then the average over all events is
taken.\footnote
{
For a survey of types of factorial \mom see \rfl \ct{msyst}.
}
These \mom are determined for a flat-shape \od distribution.  In order to
account for non-flatness, we apply the correction procedure proposed in
\rfs
\ct{pwhy,cor}, so that the corrected modified factorial \mom \ct{ks} are
given in the reduced form,

\be
F_q^C=F_q/R_q\, , \;\;\;
R_q=
\overline {N_m^{[q]}}/
{\bar N}_m^{[q]} \, ,
\la{rm}
\ee
where $R_q$ is the correction factor, and it is equal to unity for a flat
\od
distribution.\\

The non-statistical \flus have been shown \ct{bp} to lead to increasing
factorial \mom with increasing number (decreasing size) of the {\psh}
regions, or bins. Such an increase, expressed as a scaling law,

\be
F_q(M) \, \propto M^{\varphi_q}
\qquad (M\to \infty),
\qquad 0<\varphi_q\leq q-1,
\la{fi}
\ee
indicates the presence of self-similar dynamics.  This increase is called
\int and the powers $\varphi_q$, or  \int slopes, show the
strength of the effect. 
The
size of the smallest bin under investigation is limited by the
characteristic correlation length (saturation effect) and/or  by the
apparatus resolution \ct{bp}. In practice, saturation happens much earlier
because of statistical limitations (the ``empty bin effect'' \ct{eb}),
which
we here attempt to reduce by using the modified \momp. Therefore, in 
the following we use the term ``\intp'' to refer only to the rise of
the factorial \mom with decreasing  bin size. \\

\subsection{Factorial cumulant \mom and \gen \mupa \cors}
\la{gc}

To extract the \gen \mupa \corp, 
the technique of normalized factorial
cumulant \momp, or \cump, $K_q$, proposed in \rfl \ct{cum}, is used. The
\cum are
constructed from the $q$-particle cumulant correlation functions which vanish
whenever one of their arguments becomes independent of the others, so that
they measure the \gen \corp. 
The factorial \cum remove the influence of the statistical
component of the \cors in the same way as the factorial \momp.
\\

Similarly to the factorial \momp, we use the corrected modified
\cump, defined as

\be
K_q^C =
{\cal N}^q
\bar {k}_q^{(m)}/
\overline {N_m^{[q]}}.
\la{kmh}
\ee
The normalization factor, $\overline {N_m^{[q]}}$ (instead of $\bar
{N}_m^{[q]}$), comes from the correction procedure expressed in
\frm  (\ref{rm})
and takes
into
account the non-flat shape of the \od distribution. The 
$k_q^{(m)}$ factors are the unnormalized factorial cumulant \mom or
the Mueller \mom \ct{math}, and  represent \gen $q$-particle \cors where
the lower-order contributions are eliminated by subtracting the
appropriate combinations of the
unnormalized factorial \momp, $\al n_m^{[p]} \ar $, of order $p< q$
from the $q$th order one, {\eg} 

\be
k_3^{(m)}=\al n_m^{[3]}\ar - 3\,\al n_m^{[2]}\ar
\al n_m\ar\,
+ 2\, \al n_m\ar ^3.
\la{mm}
\ee

In order to find contributions from \gen \mupa \cors to the factorial
\mom we use the relations between the \mom and the \cum \ct{cumfi}, 
\nwp
\bea
F_2 &  = & K_2 + 1, \nonumber \\
F_3 & = & K_3 + 3K_2+ 1,  \la{kf} \\
F_4 & = &K_4 + 4K_3 + 3\overline {(K_2^{(m)}) ^2}+6K_2+1, 
\nonumber \\
F_5 & = & K_5+5K_4+
      10\, \overline {K_3^{(m)}K_2^{(m)}}+
      10K_3 + 15\, \overline {(K_2^{(m)}) ^2}+10K_2+1, \; {\rm \et}
\nonumber 
\eea
with $K_q ^{(m)}=k_q^{(m)}/\langle n_m\rangle ^q$.  Here and below in this
section, for brevity we omit the superscript $C$.  
These relations are exact for the ``vertical'' \mom and \cum and 
the errors introduced by using them with corrected horizontal \mom and
\cum are negligibly small, except when very high orders are considered for
variables whose \od distributions are markedly non-uniform 
\ct{rev2}.\\

The composition of the
$q$-particle dynamical \flus from the \gen lower-order $p$-particle \cors
is tested by the comparison of the $q$th order factorial \mom with the
$p$-particle contribution $F_q^{(p)}$ calculated by the above \fre
(\ref{kf}), which are truncated up to the $K_p$-terms.  The excess of $F_q$
over $F_q^{(p)}$ demonstrates the importance of \cors of order
higher than $p$ in the measured $q$-particle \flup.
\\

For example, the two-particle contribution
$F_4^{(2)}$ to the fourth-order factorial moment can be expressed as
\be
F_4^{(2)}  =  3\overline {( K_2^{(m)}) ^2} + 6K_2+1,
\la{f42}
\ee
while the contribution from two and three-particle correlations,
$F_4^{(2+3)}$, is  
\be
F_4^{(2+3)}  = 4K_3 + 3\overline {(K_2^{(m)}) ^2}+6K_2+1.\\ 
\la{f43}
\ee

\section{Experimental details}
\la{data}

The data used in this study were recorded with the \OP detector \ct{Od} at
the
LEP \ep collider at CERN.  The analysis is restricted to charged particles
measured in the central tracking chambers. The inner vertex
detector has a high precision in impact parameter reconstruction. The
large
diameter jet chamber and outer layer of longitudinal drift chambers allow
accurate measurements in the planes perpendicular and parallel to the
beam axis.  The jet chamber provides up to 159 space points per track, and
allows particle identification by measuring 
the ionisation energy loss, $d$E/$d$x, of charged particles \ct{OdE}.
All tracking chambers are surrounded by a solenoidal coil
providing a magnetic field of 0.435~T along the beam axis. 
The resolution of the component of momentum perpendicular 
to the beam axis 
is $\sigma (p_t)/p_t= \sqrt{(0.0023p_t)^2+(0.018)^2}$ for $p_t$ in
GeV/$c$, and the resolution in the
angle $\theta$ between the charged particle's direction and the electron 
beam is $\sigma(\theta)=5$ mrad 
within the acceptance of the analysis presented here.
In
multihadronic events, the ionisation 
energy loss measurement has been obtained with a resolution of
$\simeq 3.5$\% for tracks with 159 measured points. 
\\

The present study was performed with a sample of approximately
$4.1$$\times$$10^6$ hadronic {\z} decays collected from 1991 through 1995.
About 96\% of this sample was collected at the {\z} peak energy and the
remaining part was collected within $\pm 3$ GeV of the peak. Over 98\% of
charged hadrons were detected. 
\\ 

The event selection criteria are based on the multihadronic event
selection algorithms described in \rfs \ct{O,Os}, and are similar to
those used in other LEP \int studies \ct{D,A,L3}.
\\

For each event, ``good'' charged tracks were accepted if they 

\begin{itemize}
\item had at least 20 measured points in the jet chamber;
\item had a first measured point closer than 70 cm from the beam axis;
\item pass within 5 cm of the \ep collision point in the projection
      perpendicular to the beam axis, with the corresponding distance 
      along the beam axis not exceeding 40 cm;
\item had a momentum component transverse to the beam direction greater
      than 0.15 GeV/$c$;
\item had a momentum smaller than 10 GeV/$c$;
\item had a measured polar angle of the track with respect to the
      beam direction  satisfying $|\cos \theta|<0.93$;
\item had a mean energy loss, dE/dx, in the jet chamber  smaller than
      9 keV/cm to reject electrons and positrons.
\end{itemize}

Selected multihadron events were required to have

\begin{itemize}
\item at least 5 good tracks;
\item a momentum imbalance of $<0.4\sqrt{s}$, which is defined as the
      magnitude of the vector sum of the momenta of all charged particles;
\item the sum of the energies of all tracks (assumed to be pions) greater
      than 0.2$\sqrt{s}$; 
\item $|\cos \theta_S| <0.7$, where $\theta_S$ is the polar angle of the
      event sphericity axis with respect to the beam direction. The
      sphericity axis is calculated using all good tracks and
      electromagnetic and hadronic calorimeter clusters. 
\end{itemize}

The first three requirements
provide rejection of background from non-hadronic
{\z} decays, two-photon events, beam-wall interactions, and beam-gas
scattering. The last requirement ensures that the event is well contained
in the
most sensitive
volume of the detector.  A total of about $2.3$$\times$$10^6$ events
were selected for further analysis. 
\\

We also used two samples of about 2$\times$$10^6$ simulated
events each,
generated at {\z} peak energy using the following
MC generators: 
\begin{itemize}
\item \JT version 7.4 \ct{js74} with the parton shower followed
        by string formation and fragmentation,
\item \HW version 5.9 \ct{herwig} with the parton shower followed
        by cluster fragmentation.
\end{itemize}   
The parameters of both MC codes were tuned with \OP data \ct{js74o} to
provide a good description of the distributions of the measured
event-shape variables and \od  kinematic variables. 
\\

In this study we chose rapidity, azimuthal angle and transverse momentum
as individual track kinematic variables. These are frequently used in
multihadronic
studies \ct{krev} and, in particular, for  \int and correlation analyses
\ct{rev1,rev2,revd,revo,bevar}. 
To make our study compatible with other investigations carried out in \ep
annihilations, these variables are calculated with respect to the
sphericity axis as follows: 
\begin{itemize}
\item Rapidity, $y=0.5\ln [(E+p_{\|})/(E-p_{\|})]$ with $E$ and
$p_{\|}$ being the energy (assuming the pion mass) and longitudinal
momentum of the particle in
the interval $-2.0$$\leq$$y$$\leq$$2.0$.

\item Transverse momentum in the interval
$-2.4$$\leq$$\ln (\pT)$$\leq$$0.7.$
The log scale is introduced, since the exponential shape of the
$\pT$-distribution causes instability in the average multiplicity
calculations, even for $\pT$ bins of intermediate size.

\item Azimuthal angle, $\phi$, calculated with respect to the eigenvector
 of the momentum tensor having the smallest eigenvalue,
 in the plane
  perpendicular to
  the sphericity axis. The angle $\phi$ is defined in the
  interval
  $0$$\leq$$\phi$$\leq$$2\pi$. 
\end{itemize}

\noindent The \od distributions of the data sample and of the MC
(corrected to the
hadron level, see below)  are shown in \fig \ref{histo}. In the following
study the maximum number of bins is taken to be $M_{\rm max}$=400, so that
the
one-dimensional minimal bin size of the above kinematic variables are: 
$\delta y_{\rm min}$$=$$0.01$, 
$\delta \phi_{\rm min}$$=$$0.9^\circ$ 
and
$\delta (\ln \pT)_{\rm min}$$\simeq$$0.008$. 
$M_{\rm max}$ is the same as was chosen in our previous \int
publication \ct{O} and is the largest value of $M$ used so far.  The
experimental resolution of each variable was estimated by a MC simulation.
It was found that the \OP detector allows the study of an intermittency
signal down to distances of the above mentioned minimal bin sizes,
although detector effects become important for bin sizes less than 0.04
in rapidity, smaller than 3$^\circ$ in azimuthal angle and less
than 0.02 in $\ln\pT$. 
The distributions in several dimensions have the advantage that the event
phase space may be subdivided into many more  bins than $M_{\rm max}$
before the limits of detector resolution are reached.
As  in our former analysis, the smallest bin sizes used are found not to
affect the observations. 
\\

To correct the measured \mom for the effects of geometrical acceptance,
kinematic cuts, initial-state radiation, resolution and particle decays,
we apply the correction procedure  adopted in our
earlier factorial
moment study \ct{O} and in analogous investigations done with
\ep annihilation\ct{D,A,L3,291,SL}.  Two samples of multihadronic events
were generated
using the \JT 7.4 MC program. The first sample does not include 
the effects of initial-state radiation, and all particles with lifetimes
longer than 
3$\times$$10^{-10}$ seconds 
were regarded as stable.  The generator-level
factorial \mom
and \cum are calculated directly from the charged particle
distributions of this sample without any selection criteria. The second
sample was generated including the effects of finite lifetimes and
initial-state radiation and was
passed through a full simulation of the \OP detector
\ct{detsim}. 
The corresponding detector-level \mom were calculated from
this set using the same reconstruction
and selection algorithms as used for the measured data. The corrected \mom
were then
determined by multiplying the measured ones by the factor

\be
U_q(M)=X_q(M)_{gen}/X_q(M)_{det}\; ,
\la{cf}
\ee
with $X_q=F_q^C$ or $K_q^C$  defined by \fre (\ref{rm}) and (\ref{kmh}). 
\\

The correction factors $U_q$ have also been computed using the \HW event
sample. For both \JT and \HW MC generators, the correction factor tends to
be less than unity and rises with order $q$ as has been observed
earlier \ct{O,D,L3}. 
The difference between the $X_q$ quantities calculated with the \JT and 
\HW generated samples has been used in the estimation of the systematic 
\errp. 
The statistical \err on the \JT $U_q$ factor have been also incorporated
into the systematic \err in this analysis.
\\

Another contribution to the systematic \err has been evaluated by changing
the above track and event selection criteria.  The \mom have been
calculated from the data sample of about two million events with the
following
variations in the selection criteria: the first measured point was
required to be
closer than 40 cm to the beam axis, the requirement 
of the transverse momentum with respect to the beam axis was removed, 
the total momentum was required to be less than 40 GeV/$c$, the charged
track
polar angle acceptance was changed to $|\cos \theta|<0.7$, and the
requirement 
on the mean
energy loss was removed. The deviations of the \mom with these changes
modify the results by no more than a few percent and do not influence
their behaviour.
\\

The total errors have been calculated by adding the systematic and
statistical \err in quadrature. The systematic \err are shown separately
in the figures (except in those given in Section \ref{fcm:mp}) and
are dominant
at
large $M$. Statistical \err based on the MC samples are similar to
those obtained from the data.
\\

It was verified that the results do not appreciably change if one
removes from the 
analysis those events which were taken at energies off the {\z} peak
energy. 

\section{Results and discussion}
\la{res}

\subsection{The measurements}
\la{fcm:exp}

In this section we present the factorial \mom $F_q^C$ and the \cum
$K_q^C$, defined in \fre (\ref{rm}) and (\ref{kmh}), respectively, and
compute them in the $\3d$ {\phs} and its projections. 
The \mom are shown in \fgs \ref{1f} to \ref{3f} and the \cum are given in
\fgs \ref{1c} to \ref{3c} as a function of $M$, the number of bins in one,
two and three dimensions. 
Both the factorial \mom and the \cum are measured up to the fifth order.
The second-order \cum are not shown since their behaviour is determined
directly by the second-order factorial \momp, as can be seen from \fre
(\ref{kf}). 
The higher-order \cum behave differently from the same-order \mom because
the \cum contain combinations of lower-order \flup, which are taken into
account in \frm (\ref{kmh}) by means of the Mueller moments as in \frm
(\ref{mm}).
\\

Overall, the factorial \mom and the \cum depend very similarly
on the bin width. The factorial \mom obey the
scaling-law of \frm (\ref{fi}) in almost all the phase-space
projections
(except in the
$\pT$-subspace), and  the \cum show analogous intermittent behaviour up to
high orders. 
This behaviour  becomes more pronounced when the analysis is extended
to several dimensions where, in contrast to the one-dimensional case,  no
saturation with decreasing bin size is observed.
The largest \mom and \cum and the largest \int slopes are found in the
$\yf$ subspace.
\\

The saturation at small bin sizes observed in the one-dimensional analysis
in
rapidity and azimuthal angle (\fgs \ref{1f} and \ref{1c}) agrees with that
predicted by QCD \ct{revd} and is a consequence of the transition to the
regime where the running of the QCD coupling $\alpha_s$ comes into play.
The dynamics governing particle density fluctuations in small bins occurs
at low energy scales with larger values of $\alpha_s$. 
One can see that the \mom and the \cum actually
have steep linear rises at 
$M$$\lesssim$$20$ ($\delta y$$\gtrsim$$0.2$, $\delta \Phi$$\gtrsim$$18^\circ$)
and level off 
at large values of $M$.  \fgs \ref{1f} and \ref{1c} show that the
transition point shifts to larger $M$ (smaller bin sizes)  as the order
$q$ of the \mom increases, also in accordance with QCD calculations.\\

The enhancement of the \int effect in higher dimensions as shown in \fgs
\ref{2f} and \ref{3f} for the factorial \mom and in \fgs \ref{2c} and
\ref{3c} for the \cump, has been attributed to ``shadowing'' {\ie},
studies
in lower dimensions lose information due to projection
\ct{pj,3d}. A model of emission of strongly collimated particles,
clustered in both rapidity and azimuth, has been suggested in \ct{pj}. In
the framework of this so-called ``pencil-jet'' model, a strong increase of
the factorial \mom is expected in the $\yf$ subspace compared to $y$ or
$\phi$ separately.  Although formation of such jet structures has been
confirmed experimentally, the increase has been found to be much less than
that predicted \ct{rev2}.\\

Jet structure also explains the behaviour observed in the $\yp$ and $\fp$
plots of \fgs \ref{2f} and \ref{2c}. Indeed, the \mom and the \cum in
$\yp$ and $\fp$ for the same total $M$ are not found to increase
as compared
to those in $y$ and $\phi$, respectively, since there is no \int in the
transverse momentum subspace.  Similarly, the \mom in $\yf$ are
approximately equal to those in $\3d$ at the same total $M$. At the same
time, the \int  is seen for larger $M$ in higher dimensions,
indicating the presence of the dynamical \flus and \cors in additional
dimensions.
\\ 

In \rfl \ct{bevar} it was claimed that the increase of the factorial \mom
with the addition of new dimensions is a trivial consequence of a {\psh}
factor and has nothing to do with the jet formation mechanism. Our
observations show that this is true if one compares the \mom at the same
abscissa $M^{1/d}$, where $d$ is the dimension of the subspace.
However, comparing multidimensional \mom
(and the \cump) at the same {\it total} number of bins, one obtains
contributions of
self-similar structure from different projections, as is the case with the
jet-structure contributions to the $\yf$ \momp.  \\

The values of the \cum are positive in most of the cases, indicating that
\mupa dynamical \cors indeed are significantly present in the
particle-production process. Large values of the \cum of the order
$q$$\geq$4 are seen in $\yf$ and $\3d$. Non-zero high-order \cum are
also found
in  rapidity and $\yp$. 
This shows  that the  factorial moments at $q$=5, the highest
order considered here, have important contributions from
lower-order \corp, a point discussed in Section \ref{fcm:mp}. 
\\

\subsection{Comparison with the \MC models}
\la{fcm:sim}

In \fgs \ref{1f} through \ref{3c} the data are compared with the
predictions of the \JT and \HW MC models. Both MC models describe the
general behaviour of the factorial \mom and \cum and show significant
positive \mupa \corp. 
\\

The one-dimensional factorial \mom (\fig \ref{1f}) and \cum (\fig
\ref{1c})  in rapidity and in azimuthal angle show that while the MC
describe the data rather well at small $M$ (large bin sizes), the models
tend to fall below the data, starting at intermediate $M$. The
discrepancy rises with $M$ and with the order of the \mom $q$. The
saturation
effect sets in earlier in the MC models than it does in the data.
In the transverse momentum projection, the models show quite different
behaviour. \JT underestimates the \mom and the \cump, whereas
\HW strongly overestimates them. 
\fgs \ref{2f}, \ref{3f}, \ref{2c} and  \ref{3c} show that there is
a better agreement between the data and the MC in high dimensions. 
\\

From these comparisons one can conclude that both MC models reproduce the
data well while neither of them is particularly preferred. 
The perturbative parton shower, on which both MC models are based,
seems to play an important role in the origin of the dynamical
\flus and \cors in \ep annihilation.
The observed differences between the two MC descriptions indicate that the
last steps of the hadronization process are not described  correctly
\ct{rev2}. Contributions from additional mechanisms 
to the observed \flus and \cors are not excluded.

\subsection{Contributions from \mupa \cors}
\la{fcm:mp}

This section describes the contributions 
of \gen \mupa \cors to dynamical \flup.  To this end we compare in \fgs 
\ref{1fc}--\ref{3fc} the measured corrected factorial \mom
to the lower-order contributions, $F_q^{C(p)}$,
calculated using \fre (\ref{kf}).
\\

\fig \ref{1fc} shows the one-dimensional case. The \flus in transverse
momentum are not shown since they do not exhibit \int behaviour (see \fig
\ref{1f}).  A significant contribution of high-order \gen \cors appears.
Two-particle correlations in rapidity and in azimuthal angle are
insufficient to explain the measured three-particle \flup. At $q$$=$$4$,
four-particle \cors are also necessary. 
The
importance of four-particle \gen \cors is also demonstrated by the
five-particle \flus in the rapidity subspace, where the addition
of the fourth-order cumulants becomes essential.
The fifth-order moment study cannot be performed for the azimuthal angle
variable because the non-uniformity of the $\Phi$ spectrum leads to large
values of the correction factor $R_5$ which makes the relations (6)
inapplicable. 
The difference between the \mom and the correlation
contributions increases with decreasing bin size.\\

The \gen multiparticle contributions, also seen to be important in the
two-dimensional $\yf$ analysis, are shown in \fig \ref{2fc0}.  The failure
of the genuine two- and three-particle \cors to describe the four-particle
dynamical \flus indicates a significant four-particle contribution in the
observed high-order fluctuations.  The need to account for higher-order
\cors is visible also in $F_5$ for small bin-sizes. The comparison of
$F_3^C$
and $F_3^{C(p)}$ for other two-dimensional subspaces, $\yp$ and $\fp$, is
shown in \fgs \ref{2fc1} and \ref{2fc2} and also indicates a considerable
contribution from three-particle \gen \corp.  As in the one-dimensional
case in $\phi$, the large $R_q$ factor for $q$$>$$3$ does not allow use of
\fre (\ref{kf}) in these cases. The same is seen in the
three-dimensional study (\fig \ref{3fc}) for $q$$=$$5$, whereas for $q<5$,
the contribution of \mupa \cors up to the fourth order is well
illustrated.\\

\subsection{Comparison with other experiments}
\la{fcm:drex}

\subsubsection{Factorial \mom in \ep annihilation}

Studies of \int in \ep interactions have been carried out mainly in
one-dimension using the rapidity variable \ct{rev2,D,A,L3,291,SL,le}. The
first 
three-dimensional analysis of factorial \mom was perfomed by CELLO
\ct{le} in 
Lorentz-invariant \phs. \DE \ct{D}  has presented  a
three-dimensional analysis of \int   in $\3d$ {\phs} and its projections, 
as it is chosen in this paper. 
The values of the \mom and their $M$-dependence, found in all these 
studies, are similar to those obtained here.  
\\

In
one-dimensional rapidity and azimuthal angle the saturation of the
factorial \mom has been observed at the same $M$
in all investigations \ct{D,A,L3,291,SL}.  The \mom in two and
three dimensions have been found
\ct{D} to be larger and to have steeper \int slopes than those in
one dimension. Jet evolution has been suggested \ct{O,D,he} as the main
source of \mupa \flup, similarly to our finding.  However, the saturation
shown by \DE for the two-dimensional factorial \mom are not present in our
analysis due to the high statistics and the modification used to take
into account the contents of small bins. 
\\

In agreement with recent \ep studies \ct{D,A,L3,SL}, the MC
models
used here are found to describe the behaviour of the measured \momp, 
although they underestimate their magnitudes.

\subsubsection{Factorial \mom and cumulants in hadronic collisons} 

The factorial \mom have also been investigated in $\3d$ phase space and
in its projections in hadron-hadron collisons by NA22 \ct{na22}. 
\\
 
In the NA22 $\pT$-subspace analysis, hadronic interactions have
shown a visible \int effect, in contrast to its absence in \ep
annihilation.
In $\phi$, on the other hand, one finds sensitive dynamical \flus in \ep
collisions, while in hadronic collisions the \flus are strongly suppressed
by a statistical component. No saturation has been observed in $y$ and
$\pT$ subspaces in hadronic collisions.
\\

In several dimensions the factorial \mom computed in hadronic
interactions also differ from those in \ep annihialtion.  In
two-dimensions, the largest 
\mom are found to be in $\yp$, and the \int effect is
observed
to be the strongest in $\yf$. These \mom show a faster increase (larger
$\varphi_q$)  than those in one dimension, although their values are lower
and saturate already at $q$$=$$3$.  The three-dimensional \mom
show
strong increase with decreasing bin size, although their values are found
to be closer to the two-dimensional \momp.  In contrast to the linear
log-log plots of the scaling behaviour
(\ref{fi}) observed in \ep annihilation,
three-dimensional factorial \mom in hadronic collisions scale
only approximately; 
they rise more quickly than the power law of \frm (\ref{fi}).
This difference is attributed to the difference in dynamics between soft
and
hard processes that leads to isotropic dynamical \flus in \ep
annihilation and to
anisotropic ones in hadron-hadron collisions \ct{saf}.   
\\

Large cumulant values are expected in \ep annihilation due to a small
number of very
energetic particle sources,
as was mentioned in the introduction.  Indeed, the \cum measured here are
much higher than those found in hadronic interactions. 
Furthermore, they have non-zero values
for higher orders, while in hadronic collisions they are
consistent with zero at $q$$>$3.
Also, the power-law increase in $M$ is seen to be stronger in the present
data. 
\\

In hadronic interactions \ct{rev2}, both in one \ct{cumfi} and several
dimensions \ct{na22}, the factorial \mom were found to be
basically composed of two-particle \corp. 
Our study shows sensitive contributions of \cors of orders
$q$=3 and even 4 to the dynamical \flus in \ep annihilation. 
The  observed contributions increase with decreasing bin size, in
agreement with
trends seen in hadronic interactions.

\section{Summary and Conclusions}
\la{sum}

A multidimensional study of local \flus and \gen \mupa \cors in hadronic
decays of the {\z} is carried out with the four-million event  sample of
the
\OP data collected at the LEP collider.
The sphericity axis is chosen as a reference axis, and the {\phs} is 
defined by the rapidity, azimuthal angle and transverse momentum variables.
The analysis is based on the \int approach and represents  an 
investigation of  the normalized factorial \mom and, for the first time in
\ep annihilation, the normalized \cump. 
The quantities studied have been corrected to reduce the bias due to the
non-uniform shapes of the \od distributions and have been modified to
eliminate effects of finite statistics.
The factorial \mom and \cum are computed up to the fifth order and down to
very small bin sizes.
\\

The factorial \mom show an \int behaviour which is strongly enhanced as
the dimension of the subspace increases from one to three. 
In the one-dimensional analysis, the \int signal is found to be larger in
rapidity than in azimuthal angle and to vanish in transverse momentum. 
The \mom in rapidity and in azimuthal angle saturate at intermediate bin
sizes, in agreement with the QCD expectation for the transition to the
running $\alpha_s$ coupling regime. 
No saturation is observed in two and three dimensions, a consequence of
jet  formation. 
\\

Our study of the \cum shows that they have large positive values, 
indicating the existence of \gen \mupa \corp. 
The \cum are found to be much larger than those in hadronic interactions,
suggesting an increase of the \cors with the decrease of the number of
independent subprocesses present in  the reaction.
The \cum are analysed in subspaces of different dimensions, and the \gen
\mupa \cors are found to be larger in higher dimensions.
This study reveals \gen \cors up to four particles in one dimension and 
significant five-particle \cors in higher dimensions. 
Large higher-order \cors measured in the $\yf$ two-dimensional projection
confirm the jet structure of dense groups of particles.  
The \cum show \int rises which are stronger than for the corresponding
factorial \momp.
\\

The large statistics of the present analysis
allows the observation of contributions of
many-particle \cors to the measured dynamical \flup.
Considerable contributions up to four-particle  are observed in
the
expansion of the factorial \mom in terms of  \cump.  
The contributions increase with decreasing bin size reflecting the
underlying self-similar dynamics.\\

The measurements are compared to \HW 5.9 and \JT 7.4 predictions. 
In general, these \MC models are found to reproduce the behaviour of the
\mom and the \cump, while underestimating the measured values
starting at intermediate bin sizes.
High-order \mupa \cors are found to be present in both models used.  
The observations confirm jet structure formation as an important  
contribution to the \corp, but other mechanisms of the hadronization
process are possibly also relevant.
\\


\medskip
\bigskip\bigskip
\noindent
{\Large \bf Acknowledgements}
\bigskip

\noindent
We particularly wish to thank the SL Division for the efficient operation
of the LEP accelerator at all energies
 and for their continuing close cooperation with
our experimental group.  We thank our colleagues from CEA, DAPNIA/SPP,
CE-Saclay for their efforts over the years on the time-of-flight and trigger
systems which we continue to use.  In addition to the support staff at our own
institutions we are pleased to acknowledge the  \\
Department of Energy, USA, \\
National Science Foundation, USA, \\
Particle Physics and Astronomy Research Council, UK, \\
Natural Sciences and Engineering Research Council, Canada, \\
Israel Science Foundation, administered by the Israel
Academy of Science and Humanities, \\
Minerva Gesellschaft, \\
Benoziyo Center for High Energy Physics,\\
Japanese Ministry of Education, Science and Culture (the
Monbusho) and a grant under the Monbusho International
Science Research Program,\\
Japanese Society for the Promotion of Science (JSPS),\\
German Israeli Bi-national Science Foundation (GIF), \\
Bundesministerium f\"ur Bildung, Wissenschaft,
Forschung und Technologie, Germany, \\
National Research Council of Canada, \\
Research Corporation, USA,\\
Hungarian Foundation for Scientific Research, OTKA T-016660, 
T023793 and OTKA F-023259.\\


\nwp


\textheight=26.cm
\nwp

\begin{figure}
\vs{6cm}
\hs{0.5cm}
\epsfysize=12cm
\epsffile[45 150 200 500]{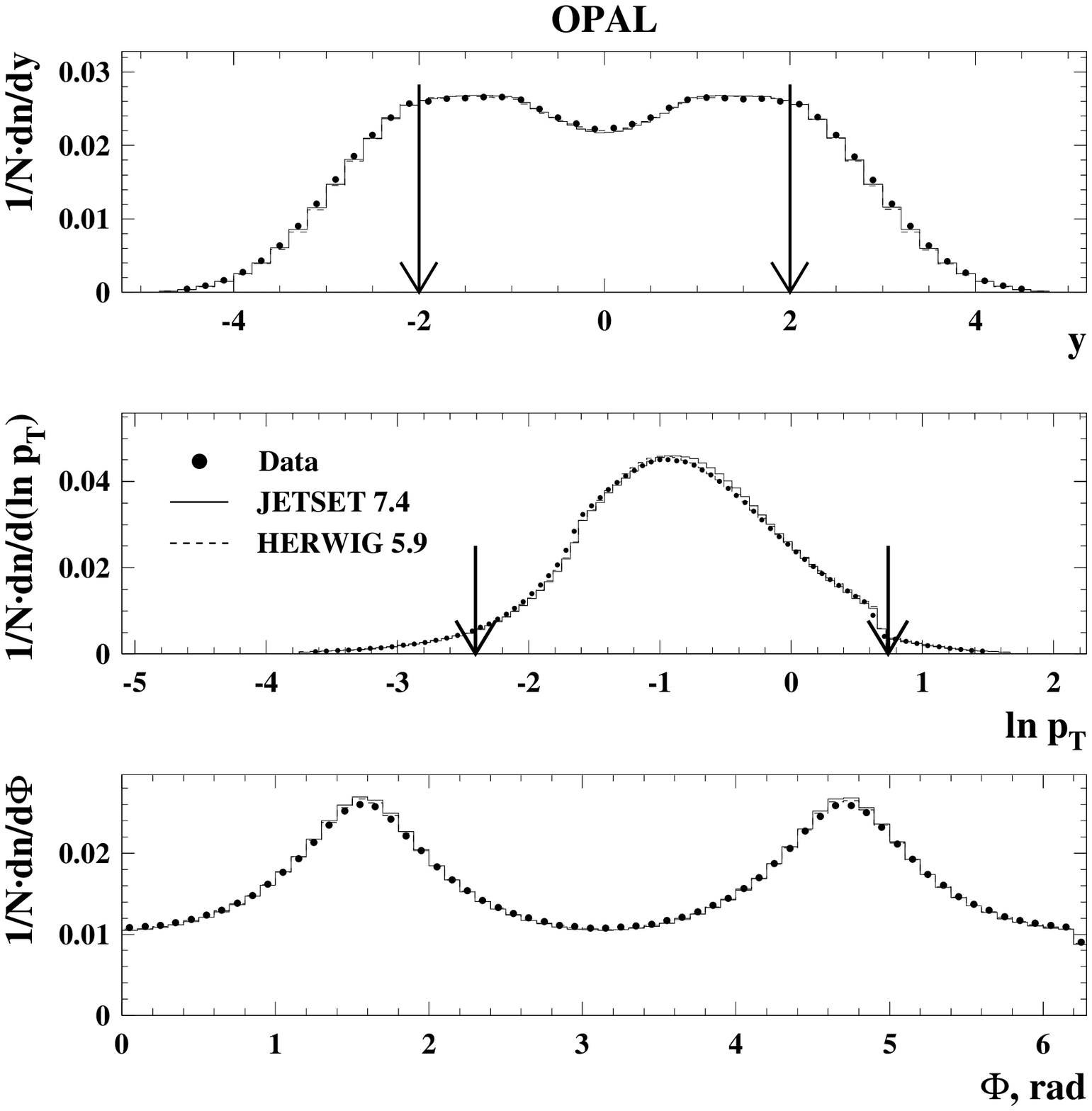}
\caption{\it 
The single-particle rapidity, the logarithm of the transverse momentum,
and azimuthal angle distributions in the data and as predicted in two \MC
models.
All kinematic variables are calculated with respect to the sphericity
axis.  The distributions are corrected for detector effects in a
bin-by-bin manner. 
The arrows show the intervals used in the analysis.}
\la{histo}
\end{figure}

\nwp
\begin{figure}
\vs{6cm}
\hs{0.5cm}
\epsfysize=12cm
\epsffile[45 150 200 500]{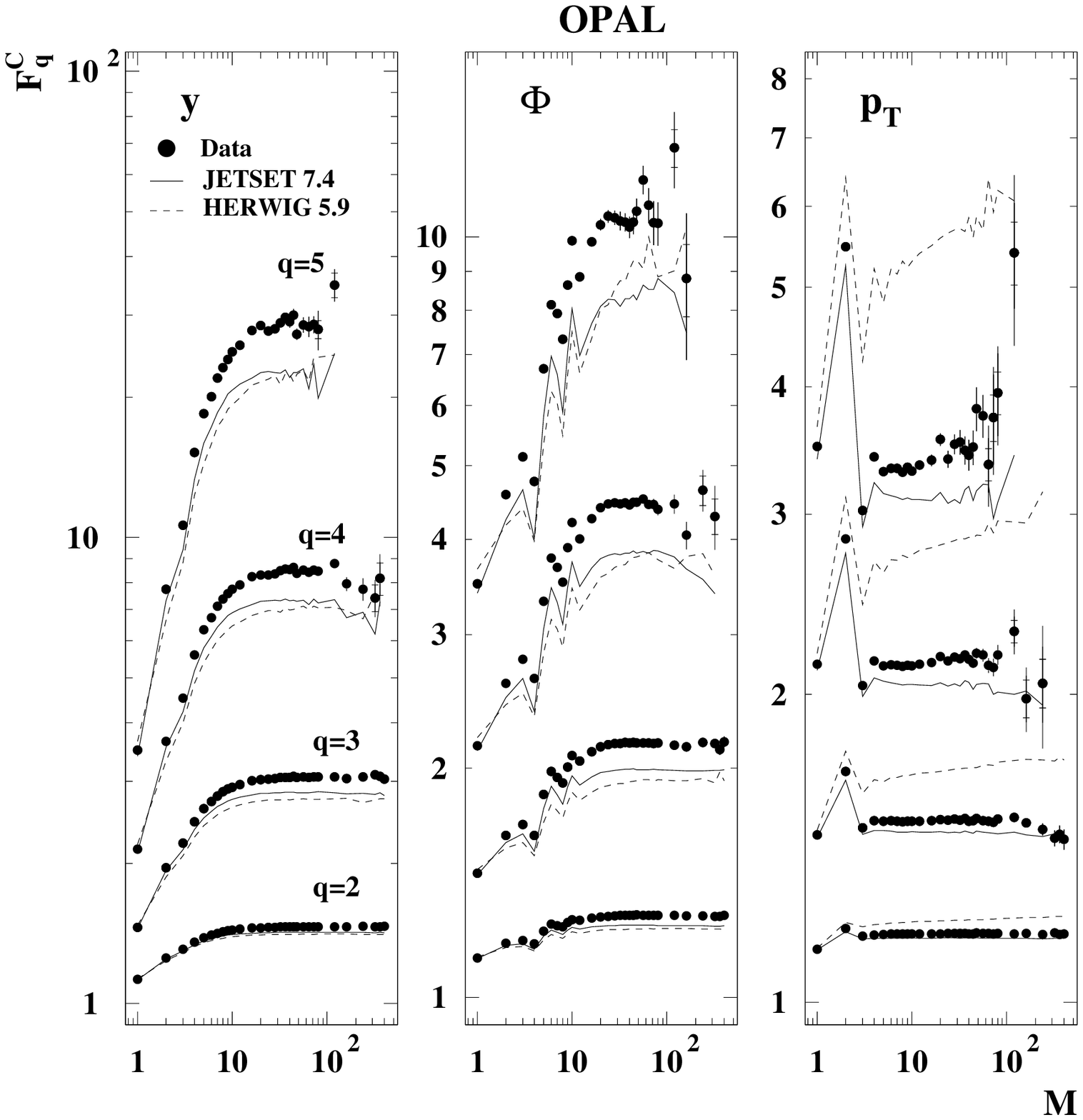}
\caption{\it 
Factorial \mom of order $q =$ 2 to 5 \vrs the number of bins $M$ of
the one-dimensional rapidity, azimuthal angle, and transverse momentum  
subspaces 
{\mode}
\lides}
\la{1f}
\end{figure}

\nwp
\begin{figure}
\vs{6cm}
\hs{0.5cm}
\epsfysize=12cm
\epsffile[45 150 200 500]{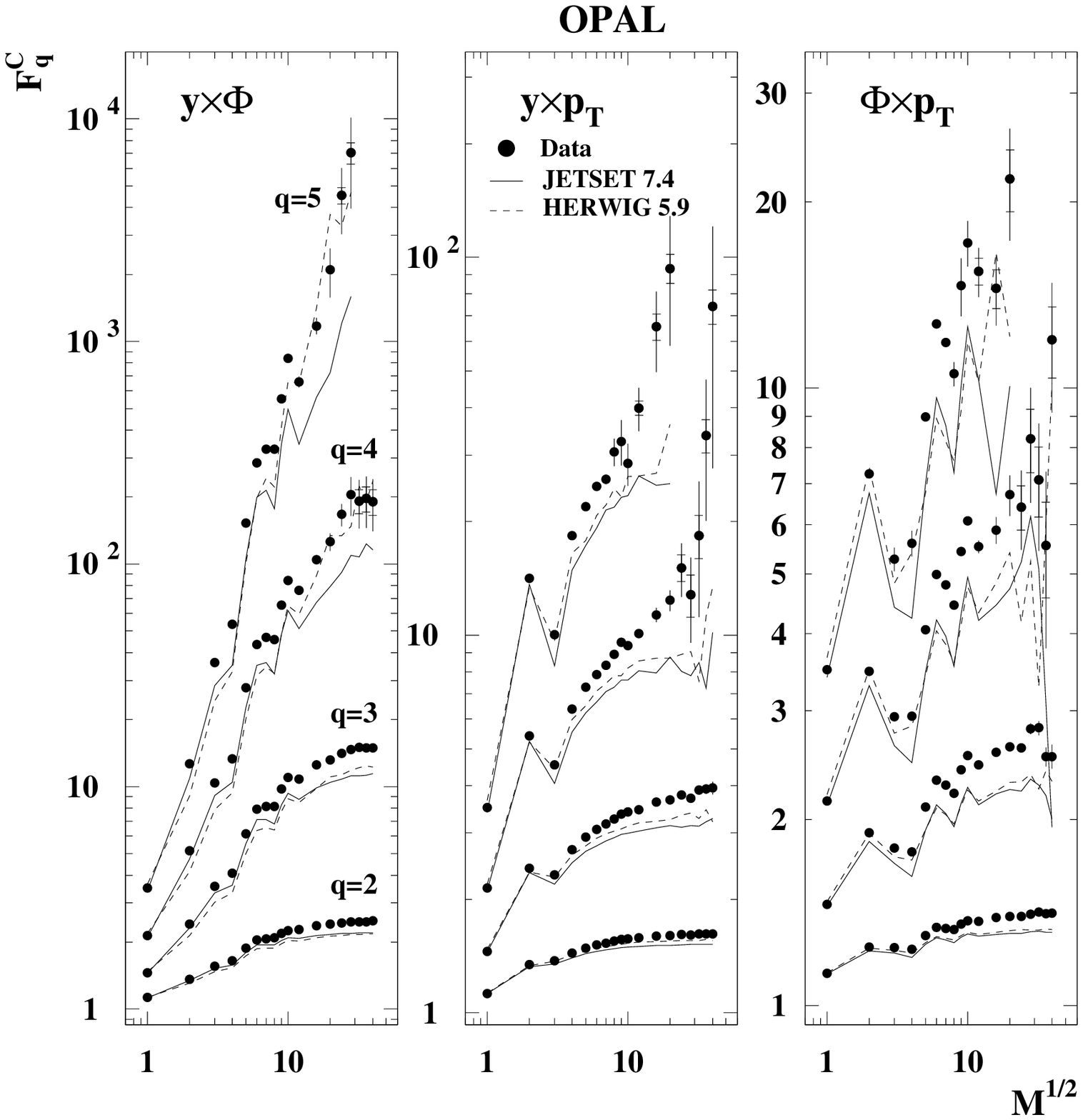}
\caption{\it 
Factorial \mom of order $q =$ 2 to 5 \vrs $\sqrt{M}$, where $M$
is the number of bins of the two-dimensional subspaces of rapidity,
azimuthal angle, and transverse momentum combinations, 
{\mode} 
\lides}
\la{2f}
\end{figure}

\nwp
\begin{figure}
\vs{6cm}
\epsfysize=12cm
\epsffile[-90 150 200 500]{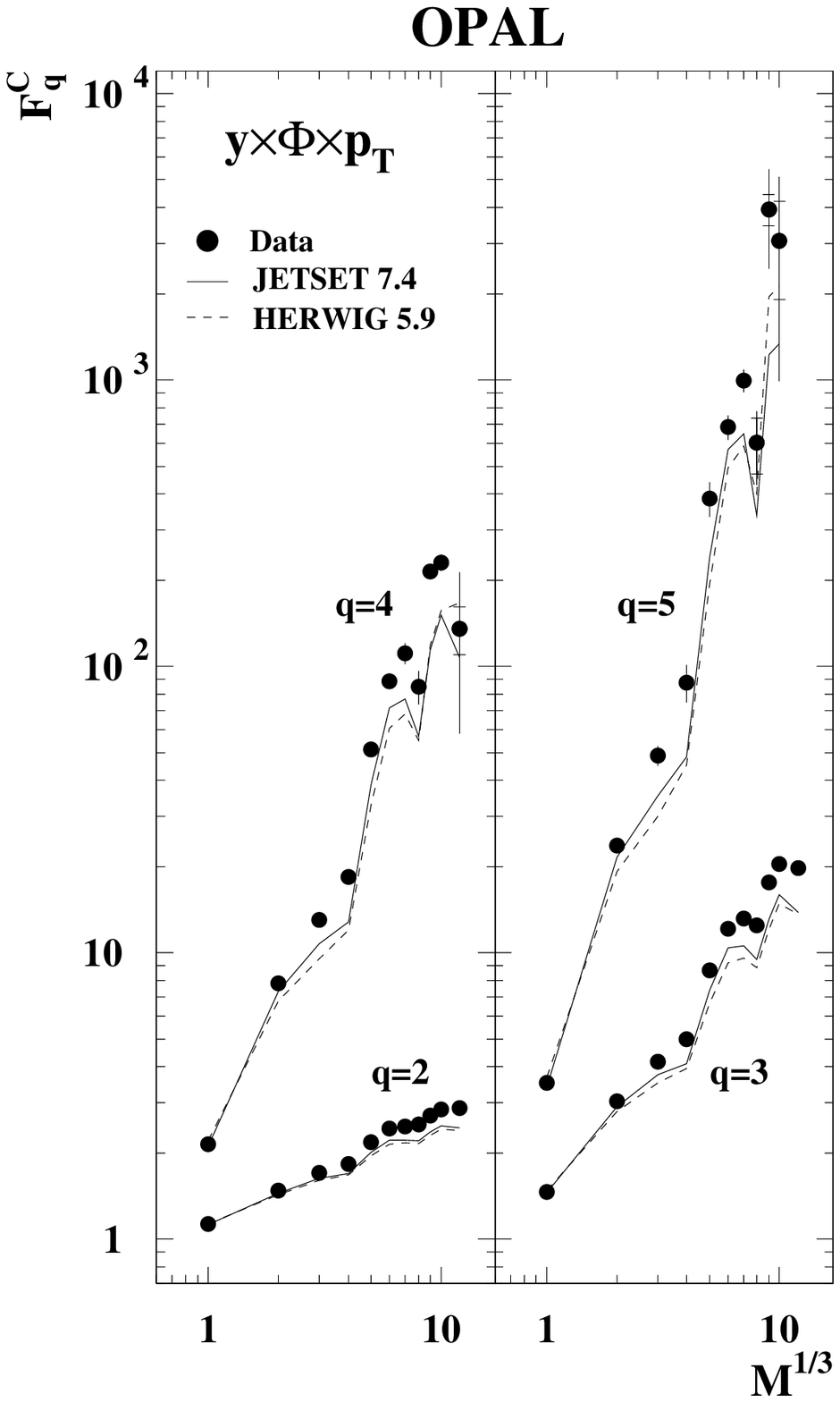}
\caption{\it 
Factorial \mom of order $q =$ 2 to 5 \vrs $\sqrt[3]{M}$, where $M$
is the number of bins of  the three-dimensional rapidity, azimuthal angle,
and transverse momentum \phs, 
{\mode}
\lides}
\la{3f}
\end{figure}

\nwp
\begin{figure}
\vs{6cm}
\hs{0.5cm}
\epsfysize=12cm
\epsffile[45 150 200 500]{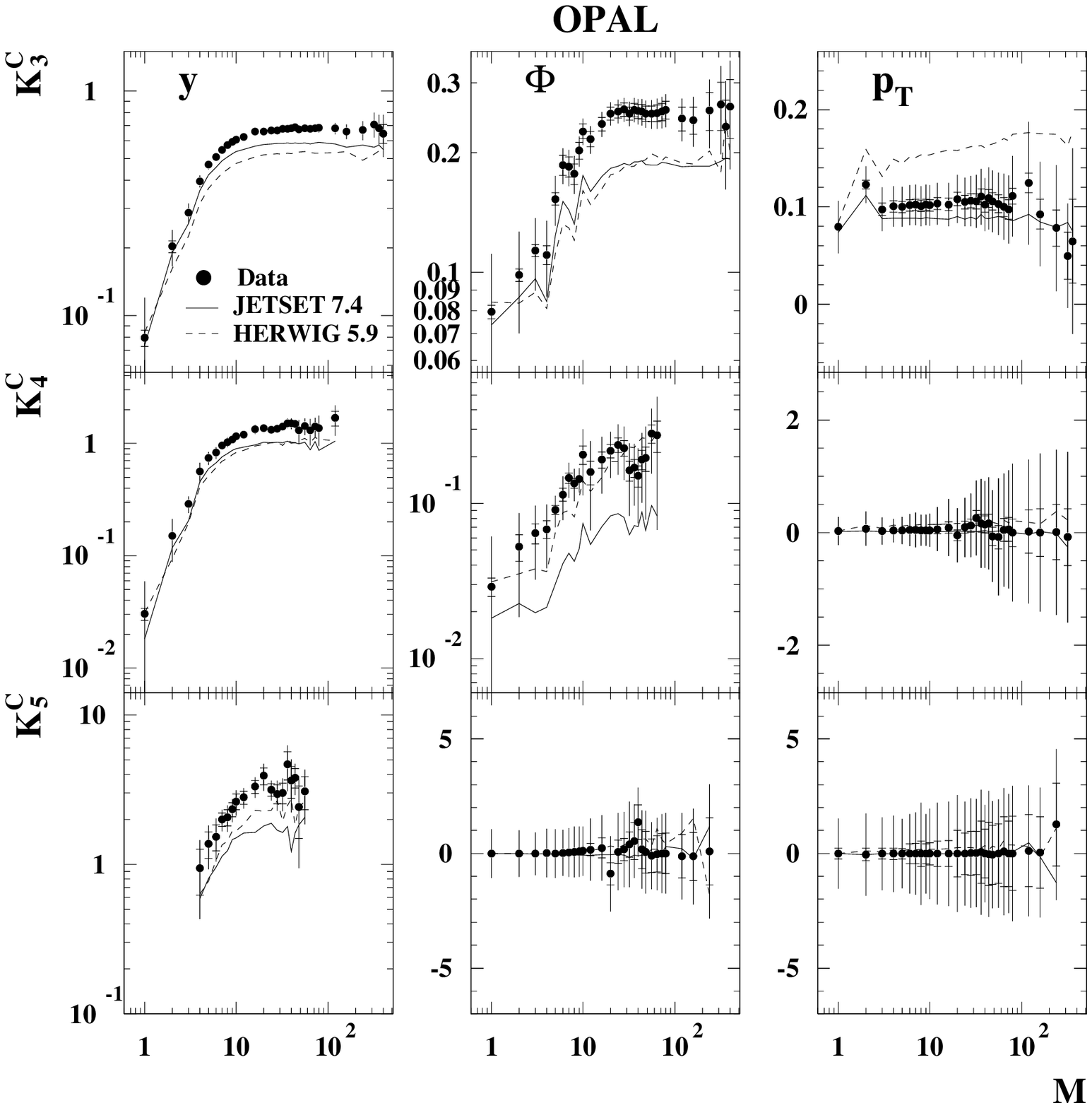}
\caption{\it 
Cumulants of order $q =$ 3 to 5 \vrs the number of bins $M$ of the
one-dimensional rapidity, azimuthal angle, and transverse momentum 
subspaces 
{\mode}
\lides}
\la{1c}
\end{figure}

\nwp
\begin{figure}
\vs{6cm}
\hs{0.5cm}
\epsfysize=12cm
\epsffile[45 150 200 500]{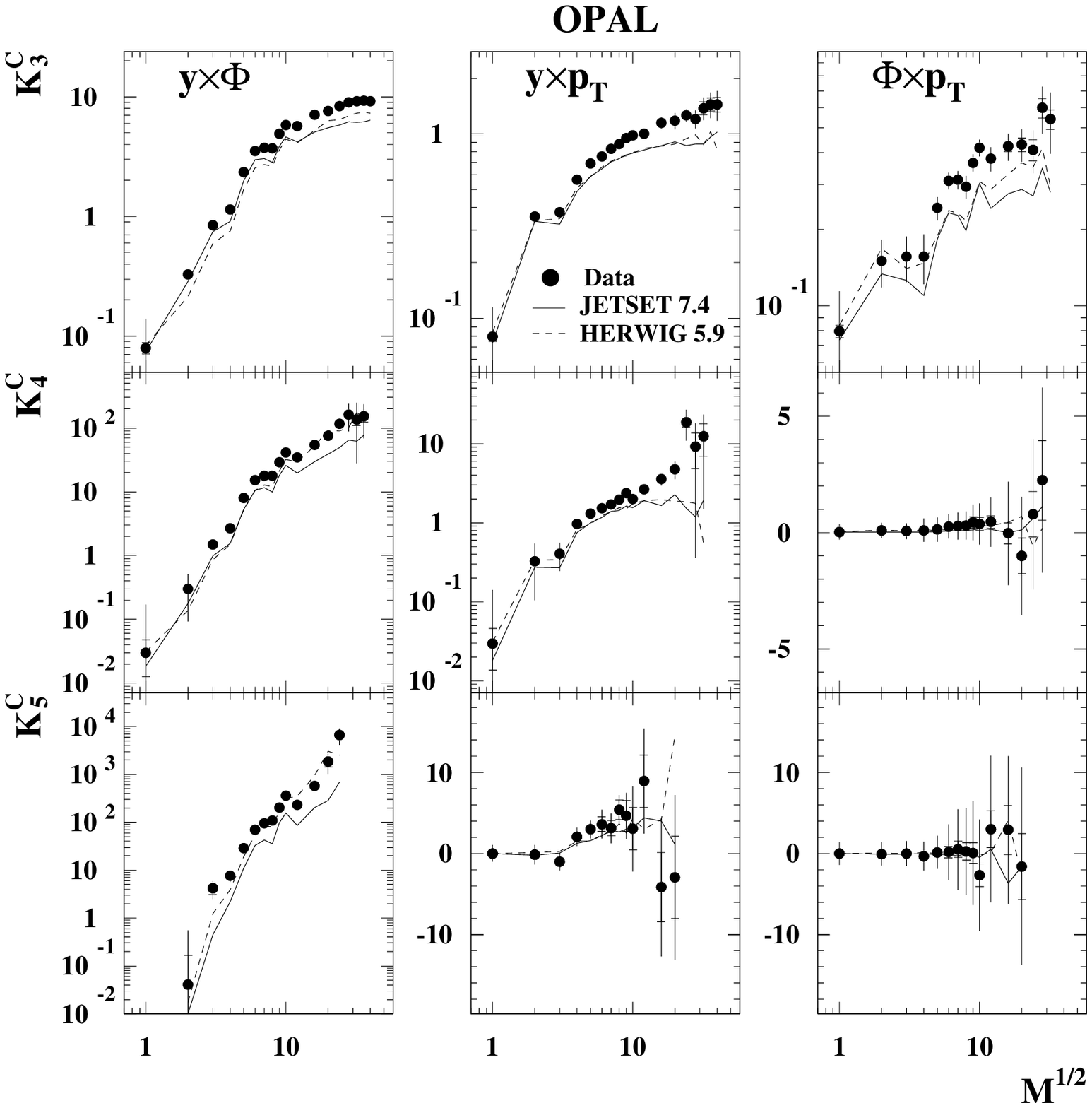}
\caption{\it 
Cumulants of order $q =$ 3 to 5 \vrs $\sqrt{M}$, where $M$ is the
number of bins of the two-dimensional subspaces of rapidity, azimuthal
angle, and transverse momentum combinations, 
{\mode}
\lides}
\la{2c}
\end{figure}

\nwp

\begin{figure}
\vs{3.cm}
\hs{-2.3cm}
\epsfysize=11.5cm
\epsffile[-50 150 200 500]{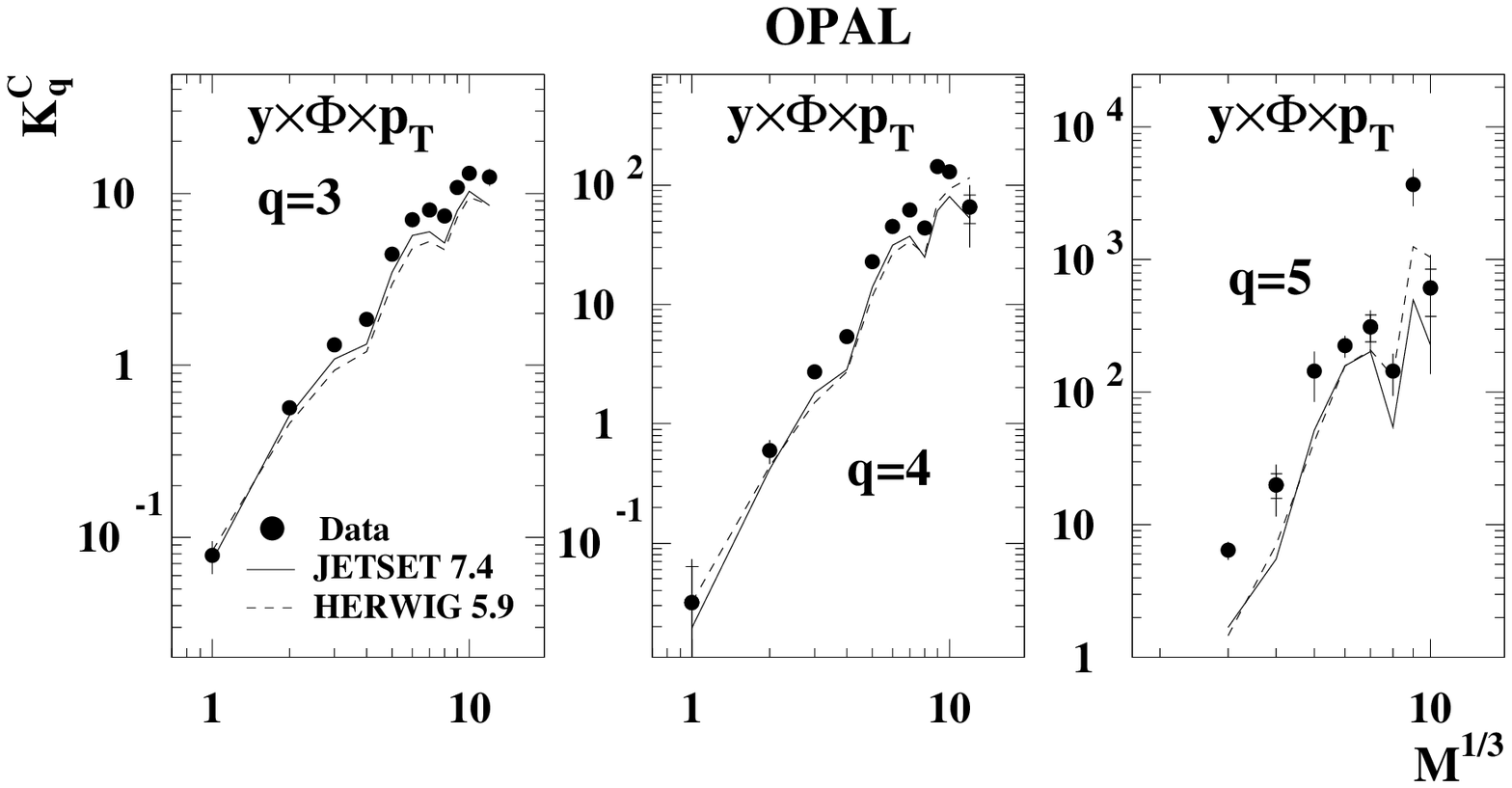}
\vs{-7.cm}
\caption{\it 
Cumulants of order $q =$ 3 to 5 \vrs $\sqrt[3]{M}$ where $M$ is
the number of bins of  the three-dimensional rapidity, azimuthal
angle, and transverse momentum \phs, 
{\mode}
\lides}
\la{3c}
\end{figure}

\nwp
\begin{figure}
\vs{6cm}
\hs{0.5cm}
\epsfysize=12cm
\epsffile[45 150 200 500]{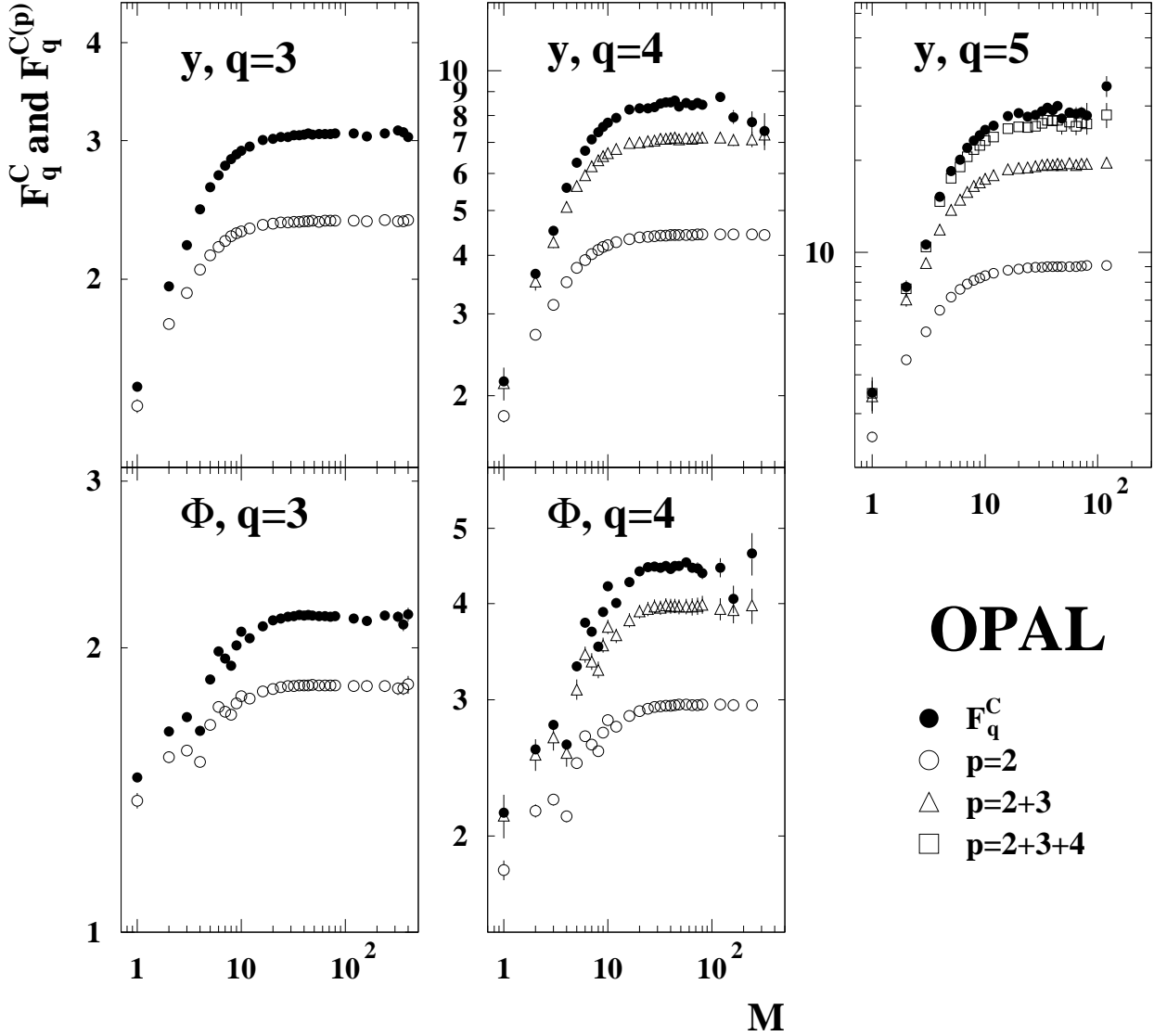}
\caption{\it 
Decomposition of the factorial \mom $F_q^C$ into multiparticle
correlation contributions $F_q^{C(p)}$ in the one-dimensional rapidity and
azimuthal angle subspaces. 
\vliet}
\la{1fc}
\end{figure}

\nwp
\voffset=-2.5cm
\begin{figure}
\vs{6cm}
\hs{0.5cm}
\epsfysize=11.5cm
\epsffile[45 150 200 500]{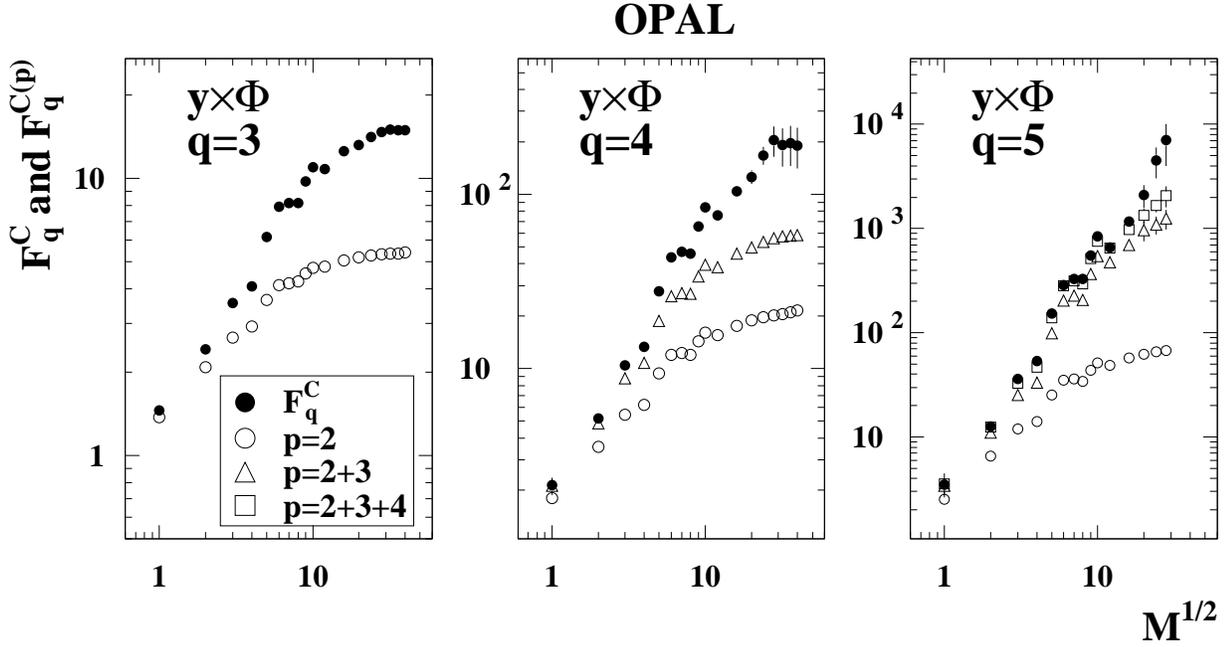}
\vs{-8.cm}
\caption{\it 
Decomposition of the factorial \mom $F_q^C$ into multiparticle
correlation contributions $F_q^{C(p)}$ in the two-dimensional subspace
of rapidity and azimuthal angle.
\vliet}
\la{2fc0}
\end{figure}

\begin{figure}
\vs{4.cm}
\hs{-2.7cm}
\epsfysize=11.5cm
\epsffile[-50 150 200 500]{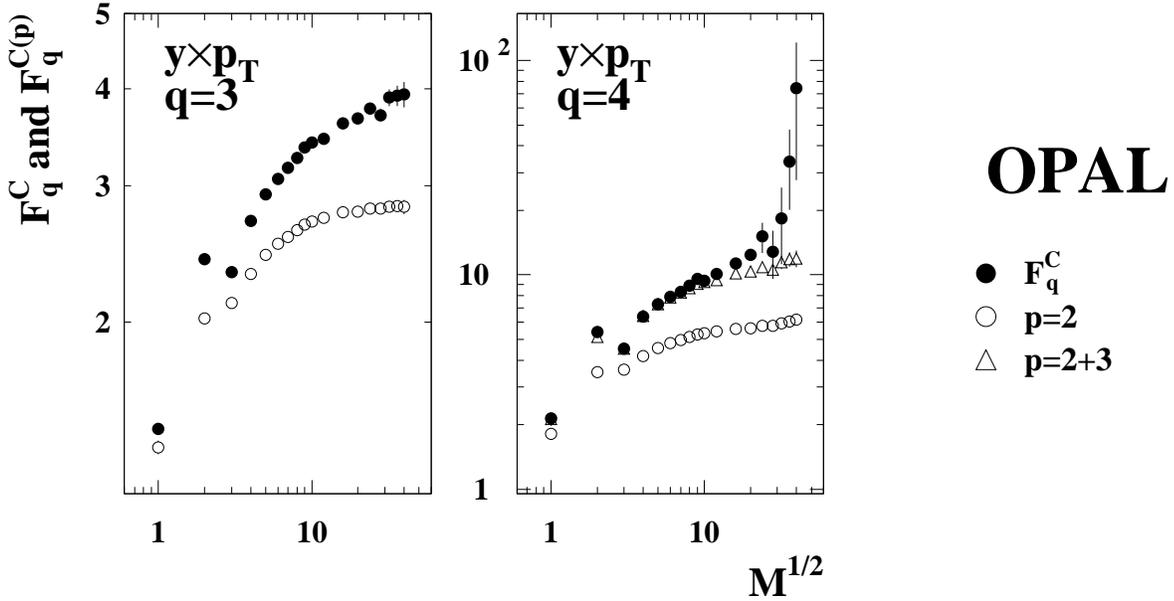}
\vspace{-8.1cm}
\caption{
\it 
Decomposition of the factorial \mom $F_q^C$ into multiparticle
correlation contributions $F_q^{C(p)}$ in the two-dimensional subspace of
rapidity and transverse momentum.
\vliet}
\la{2fc1}
\end{figure}

\nwp
\voffset=-2.5cm
\begin{figure}
\vs{6cm}
\hs{-2.cm}
\epsfysize=11.5cm
\epsffile[-50 150 200 500]{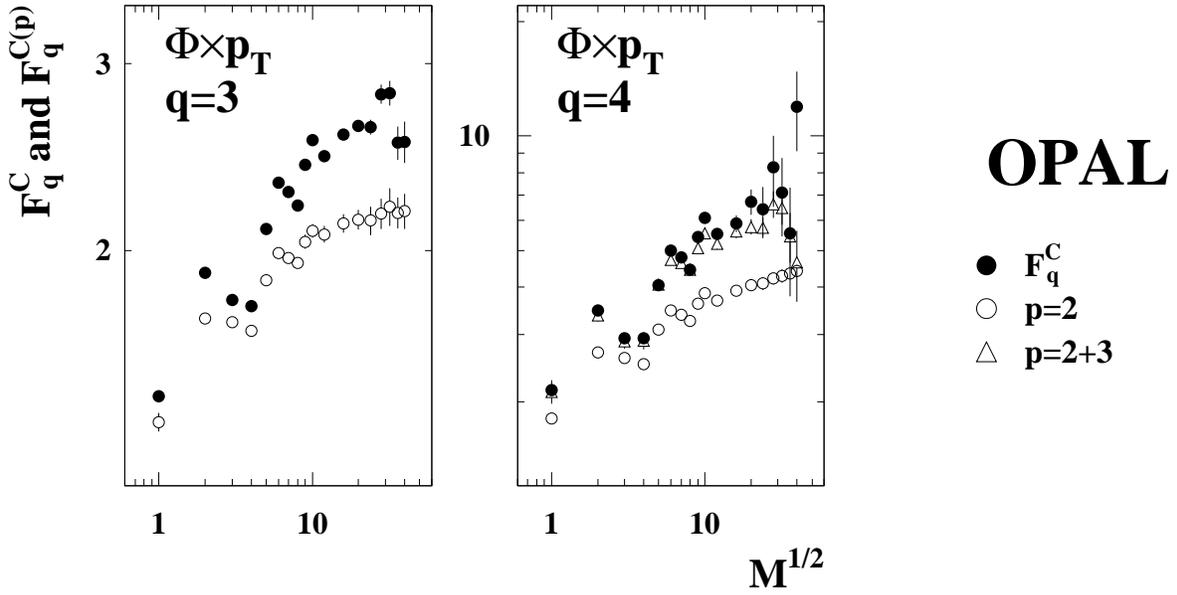}
\vs{-8.cm}
\caption{\it 
Decomposition of the 
factorial \mom $F_q^C$ into multiparticle correlation contributions
$F_q^{C(p)}$ in the two-dimensional subspace of azimuthal angle and
transverse momentum.
\vliet}
\la{2fc2}
\end{figure}

\begin{figure}
\vs{5cm}
\hs{-2.cm}
\epsfysize=11.5cm
\epsffile[-50 150 200 500]{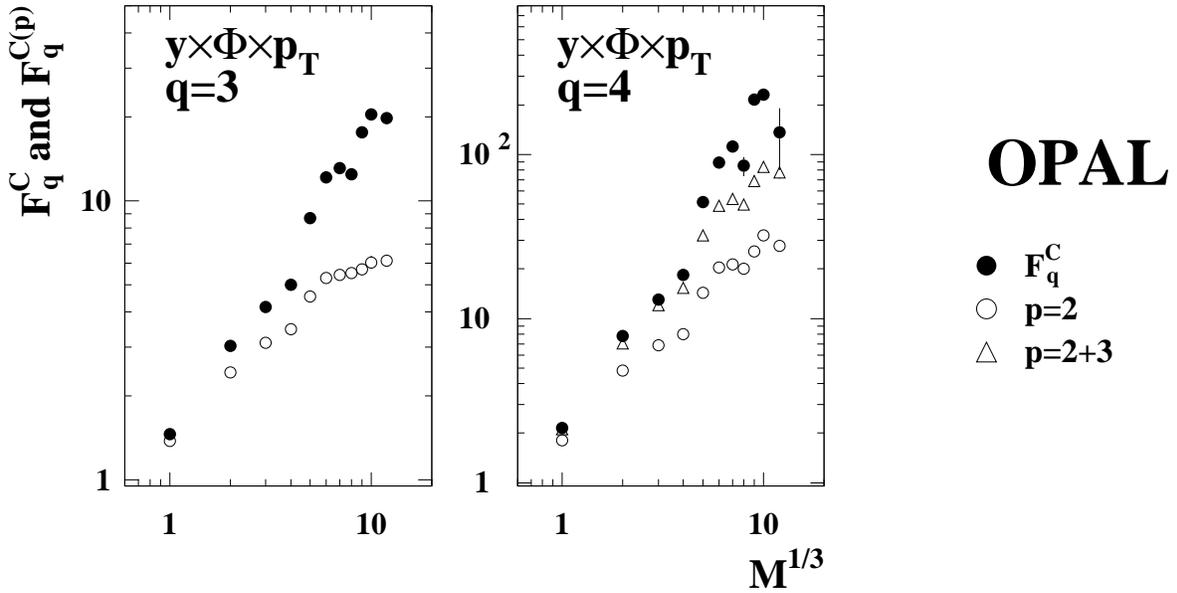}
\vs{-8.cm}
\caption{\it 
Decomposition of the factorial \mom $F_q^C$ into multiparticle
correlation contributions $F_q^{C(p)}$ in the three dimensions.
\vliet}
\la{3fc}
\end{figure}

\end{document}